\documentclass[a4paper,aps,prl,twocolumn,superscriptaddress]{revtex4}
\usepackage[usenames,dvipsnames]{xcolor}
\usepackage{graphicx}
\usepackage{amsmath}
\usepackage{amssymb}
\usepackage{esint}
\usepackage{amsfonts}
\usepackage{natbib}
\usepackage{psfrag}
\usepackage{wasysym}
\usepackage{float}
\usepackage{epstopdf}
\usepackage{color}
\usepackage[normalem]{ulem}
\usepackage{textcomp}
\usepackage{bm}
\usepackage{hyperref}
\hypersetup{linkbordercolor=blue}

\renewcommand{\d}[2]{\frac{\id #1}{\id #2}} 

\newcommand{\id}{\mathrm{d}} 

\newcommand{\e}{\mathrm{e}}

\begin{document}

\title{Hydrogel menisci: Shape, interaction, and instability}

\author{Anupam Pandey}
\affiliation{Physics of Fluids Group, Faculty of Science and Technology, University of Twente, P.O. Box 217, 7500 AE Enschede, The Netherlands}

\author{Charlotte L. Nawijn}
\affiliation{Physics of Fluids Group, Faculty of Science and Technology, University of Twente, P.O. Box 217, 7500 AE Enschede, The Netherlands}

\author{Jacco H. Snoeijer}
\affiliation{Physics of Fluids Group, Faculty of Science and Technology, University of Twente, P.O. Box 217, 7500 AE Enschede, The Netherlands}

\date{\today}

\begin{abstract}
The interface of a soft hydrogel is easily deformed when it is in contact with particles, droplets or cells. Here we compute the intricate shapes of hydrogel menisci due to the indentation of point particles. The analysis is based on a free energy formulation, by which we also assess the interaction laws between neighbouring particles on hydrogel interfaces, similar to the ``Cheerios effect". It is shown how the meniscus formed around the particles results from a competition between surface tension, elasticity and hydrostatic pressure inside the gel. We provide a detailed overview of the various scaling laws, which are governed by a characteristic shear modulus $G^*=\sqrt{\gamma\rho g}$ that is based on surface tension $\gamma$ and gravity $\rho g$. Stiffer materials exhibit a solid-like response while softer materials are more liquid-like. The importance of $G^*$ is further illustrated by examining the Rayleigh-Taylor instability of soft hydrogels.
\end{abstract}


\maketitle

\section{Introduction}

Hydrogels, a mixture of polymer network and water, constitute the extracellular matrix of animal bodies and are found in mucus, cartilage, and cornea~\cite{LAI2009}. Even though soft, these materials are tough, and perform remarkable functions such as sensing~\cite{ULIJN2007}, self-healing~\cite{Krogsgaard13}, lubricating joints~\cite{Crockett2009}, and selective filtering~\cite{LIELEG2011}. In recent years synthetic hydrogels with tailored polymer structures have been developed to serve as stimuli responsive valves in microfluidics~\cite{Beebe00}, scaffolds in tissue engineering~\cite{LEE2001}, and vehicles for drug delivery~\cite{QIU2001}. In a number of these applications liquid drops, solid particles or biological cells reside on a hydrogel interface and deform it by applying traction. This deformation induces an interaction that leads to biomechanosensing in living cells~\cite{SchwartzPRL2002,Reinhart:BPJ2008, Guo:BPJ2006}, and self-assembly of particles~\cite{Chakrabarti2014,Maha2015}. Understanding the mechanics of hydrogel interfaces is thus key in a broad variety of contexts.




\begin{figure}[t]
\includegraphics[width=85mm]{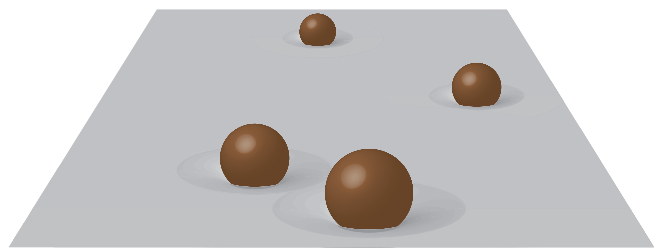}
\caption{A schematic of rigid particles on a soft elastic substrate. Each particle is associated with a deformation field around it, leading to mutual interactions.}
\vspace{-4mm}
\label{fig1}
\end{figure}

The challenge is that hydrogels have attributes of both solids and liquids. While the polymer network gives rise to an elastic (shear) modulus $G$, the interface also possesses a surface free energy $\gamma$ that plays a crucial role in the deformation and stability of these solids~\cite{mora10,paretkar14,StyleRev17}. As such, a soft elastic solid forms a ``meniscus" whenever brought into contact with a rigid object~\cite{Style13_1,Hui2016,Karpitschka16_1,ChakrabartiArxiv2018}. For solid particles on extremely soft hydrogels, with shear moduli down to $10$ Pa, these menisci indeed give rise to particle interactions~\cite{Chakrabarti2014,Maha2015} that resemble the ``Cheerios effect" -- the clumping of floating paperclips and cereals induced by liquid menisci~\cite{Vella2005,botto2012,Ershov13,loudet2011,soligno2016}. Similar interactions are found for droplets on elastomeric interfaces \cite{Karpitschka2016, pandey17}. In this context it is of particular importance to know the detailed shape of the meniscus, since ultimately this determines the nature of the interaction. Pushing the analogy with liquid interfaces, it was argued that the deformations are exponentially screened by hydrostatic pressure inside the gel \cite{Chakrabarti2014,Maha2015}. However, even though hydrostatic effects were demonstrated in the context of the Rayleigh-Taylor instability of hydrogels~\cite{Mora14}, its effect on the meniscus shapes remains to be analysed in detail.  
 
In this paper, we analyse hydrogel menisci based on a free energy formulation. As such, we are able to investigate the combined effects of surface tension, bulk elasticity and hydrostatic pressure inside the gel. We reveal the emergence of intricate meniscus shapes and quantify the particle-particle interactions that results from these. We provide a detailed overview of the regimes and scaling laws, which are governed by a characteristic shear modulus $G^*=\sqrt{\gamma\rho g}$ based on surface tension $\gamma$ and gravity $\rho g$. Stiffer materials exhibit a solid-like response while softer materials are more liquid-like. Finally, we discuss the gravity-driven instability of the hydrogel meniscus, showing how $G^*$ governs the transition from liquid to solid response.

\section{Formulation}


We start out by setting up a formalism for particles on a soft interface, from which we compute the deformations and particle interactions (Fig.~\ref{fig1}). Particles are treated as being point-like so that the theory describes the far-field behaviour, at distances larger than the typical particle size. Particle $i$ is at a position $\mathbf{x}_i=(x_i,y_i)$ and has a weight $w_i= m_i g$. Each particle resides at the interface of shape $h(\mathbf{x})$, so the corresponding gravitational energy is $w_i h (\mathbf{x}_i)$. However, there is also an energetic cost associated to deforming the interface, described by a functional $E_{\rm int}[h]$. Combined, this gives the total energy

\begin{equation}
E\left[h; \mathbf{x}_i\right] = E_{\rm int}[h] + \sum_ i w_i h(\mathbf{x}_i).
\label{e_functl}
\end{equation}
This is in the form of a simple field theory for particles of ``charge" $w_i$, coupled to the field $h(\mathbf{x})$~\cite{landau1975classical}. The degrees of freedom are therefore the discrete particle positions $\mathbf{x}_i$ and the continuous field $h(\mathbf{x})$. The field equation that describes the shape of the interface is obtained by the functional derivative of $E$ with respect to $h(\mathbf{x})$, 

\begin{equation}
\frac{\delta E_{\rm int}}{\delta h(\mathbf{x})}  + \sum_i w_i \delta\left(\mathbf{x} - \mathbf{x}_i \right)  =0,
\end{equation}
or in compact form

\begin{equation}
\sigma(\mathbf{x}) = - \varrho(\mathbf{x}),
\label{eq_eqn}
\end{equation}
where we define the normal stress $\sigma = \frac{\delta E_{\rm int}}{\delta h(\mathbf{x)}}$ and the weight distribution $\varrho(\mathbf{x}) =  \sum_i w_i \delta\left( \mathbf{x} - \mathbf{x}_i \right)$. The force $\mathbf F_i$ on particle $i$ follows as 


\begin{equation}
\mathbf{F}_i = - \frac{\partial E}{\partial \mathbf{x}_i} = - w_i \nabla h,
\label{f_int}
\end{equation}
where the gradient is evaluated at $\mathbf{x} = \mathbf{x}_i$. Here the interface shape $h(\mathbf x)$ plays a role similar to the electrostatic or gravitational potential, since its gradient gives the force. Once the interface functional $E_{\rm int}[h]$ is specified,~\eqref{eq_eqn} and~\eqref{f_int} fully define the problem.

\subsection{Interface functionals}

When deforming a hydrogel interface, there is an energy cost due its bulk elasticity, its bulk gravitational energy, as well as its surface free energy. We first briefly recap the well-known case of a liquid interface, which will provide expressions for the capillary and gravitational energies. Then we turn to the hydrogel menisci by incorporating the elastic energy. 

\subsubsection{The liquid interface}

A liquid interface is governed by capillarity and gravity, which can be described by the interface functional

\begin{equation}
E_{\rm liq}[h] = \int \id\mathbf{x} \left( \frac{1}{2}\gamma |\nabla h|^2+ \frac{1}{2}\rho g h^2  \right).
\label{E_liq}
\end{equation}
The first term is the excess surface energy due to the deformation, in the approximation where $|\nabla h|$ remains small. This is a natural approximation in the present context based on point particles, which implicitly implies a far-field description where deformations are small. The second term is the gravitational energy of an incompressible medium integrated over the vertical direction. Taking the functional derivative with respect to $h(x)$ we find the field equation

\begin{equation}
-\gamma \nabla^2 h + \rho g h = -\varrho.
\label{eq_liq}
\end{equation}
In the absence of particles ($\varrho=0$), we recover the classical Young-Laplace equation for a meniscus, where the Laplace pressure balances with the hydrostatic pressure. The typical scale of a liquid meniscus is set by the ratio $\ell_c = (\gamma/\rho g)^{1/2}$, where we defined the capillary length $\ell_c$. 

When particles are present ($\varrho\neq 0$), the meniscus will be perturbed with respect to its flat state and induce a nontrivial field $h(\mathbf x)$. In fact, if we ignore the hydrostatic term in (\ref{eq_liq}), the equation is strictly identical to Gauss's law of electrostatics (or Newtonian gravity), where $\varrho$ is the distribution of charge (or mass). These analogies have indeed been successfully exploited for particles at liquid interfaces \cite{Vella2005,dietrich2011}. However, the present formulation based on (\ref{e_functl}) has the merit that it can be extended to more general $E_{\rm int}[h]$, as is necessary for hydrogels. 

\subsubsection{The hydrogel interface}

Along with capillary and gravitational energy, any deformation of a hydrogel surface leads to strain energy inside the bulk. The poroelastic nature of hydrogels gives rise to an intricate time-dependent evolution when it is indented~\cite{Hu2010, Chan2012}. However, here we focus on the equilibrium response of the gel, which is purely elastic and can be captured by a free energy. Here we assume that the hydrogel layer is infinitely thick, and the material is incompressible (Poisson ratio $\nu=1/2$). For this case the energy based on linear elasticity can be written in explicit form (see Appendix A)

\begin{equation}
E_{\rm el} = \frac{G}{2\pi} \int \id\mathbf{x} \, \int \id\mathbf{x}' \, \frac{\nabla h(\mathbf{x})\nabla h(\mathbf{x}')}{|\mathbf{x}-\mathbf{x}'|}.
\label{E_el}
\end{equation}
The corresponding functional derivative gives the elastic stress $\sigma_{\rm el}(\mathbf x)=\frac{\delta E_{\rm el}}{\delta h(\mathbf x)}$, which after integrating by parts takes the form

\begin{eqnarray}
\sigma_{\rm el}(\mathbf x) &=&-\frac{G}{\pi} \int \id\mathbf{x}' \, \frac{\nabla^2 h(\mathbf{x}')}{|\mathbf{x}-\mathbf{x}'|}\nonumber\\ 
&=& -\frac{G}{\pi} \, \left(K\circ\nabla^2 h \right)(\mathbf x).
\label{stress_el}
\end{eqnarray}
Here we have introduced the convolution with a Green's function $K(\mathbf x) =1/|\mathbf{x}|$. 

A subtle point is that (\ref{E_el}) is derived for an elastic medium without any bulk force, i.e. without the gel's gravitional energy. For incompressible media however, the addition of a bulk force that derives from a potential can be absorbed into the hydrostatic pressure. As a consequence, the resulting bulk strains -- and therefore the resulting elastic energies -- are totally unaffected by the presence of a bulk force. Therefore, the total functional for a hydrogel interface is obtained by a simple addition

\begin{equation}\label{eq:additive}
E_{\rm int}[h] = E_{\rm liq}[h] + E_{\rm el}[h].
\end{equation}
The validity of this approach will be demonstrated explicitly at the end of the paper. Now, we immediately obtain $\sigma =\sigma_{\rm liq}+\sigma_{\rm el}$, so that (\ref{eq_liq},\ref{stress_el}) gives the field equation 

\begin{equation}
- \gamma \nabla^2 h + \rho g h  - \frac{G}{\pi} \, \left(K \circ \nabla^2 h\right)
= - \varrho. 
\label{gov_eq}
\end{equation} 
For a given distribution of particles $\varrho(\mathbf x)$, this equation can be solved analytically using integral transformation methods.

\section{Hydrogel meniscus}

\subsection{Solution}

Here we first determine the shape of the hydrogel interface indented by a single particle of weight $mg$. Owing to the linearity of the problem, the deformation due to a distribution of particles as in Fig.~\ref{fig1} is simply obtained by superposition. For the single particle, we introduce cylindrical coordinates with the particle located at the origin, so that $r = |\mathbf x|$. We find the resulting axisymmetric deformation $h(\mathbf{x})=h(r)$ by solving \eqref{gov_eq} using the Hankel transform, defined as $\hat{h}(s)=\int_0^{\infty}h(r)r J_0(sr)\id r$. Here $J_0$ is the zeroth order Bessel function. Transformation of~\eqref{gov_eq} gives

\begin{equation}
\gamma s^2\hat{h}(s)+\rho g\hat{h}(s)+{G\over\pi}2\pi s^2\hat{K}(s)\hat{h}(s)=-\hat{\varrho}(s),
\label{gov_eq_trans}
\end{equation}
where we used properties of the Hankel transform that resemble those of the Fourier transform. Namely, the Hankel transform of the Laplacian reads $-s^2\hat{h}(s)$, while the convolution $K \circ \nabla^2 h$ transforms to $-2\pi s^2 \hat{K}(s) \hat{h}(s)$. Furthermore, the Green's function $\hat{K}(s)=1/s$, while for a single point particle the weight distribution $\hat{\varrho}(s) = mg/2\pi$~\cite{soutas2012}. With this, ~\eqref{gov_eq_trans} gives a closed form solution  

\begin{equation}
\hat{h}(s)={-mg/2\pi\over\gamma s^2+\rho g+2G s}.
\label{hs}
\end{equation} 
The backward transform gives the deformation in real space, $h(r)$, which in general is done numerically. 

\subsection{Shape}

Before presenting the actual shapes of hydrogel menisci, it is instructive to discuss the length scales for $r$ that are implied by the terms in the denominator of (\ref{hs}). We remind that $s$ is an inverse length, so that the ratios of the terms give rise to two length scales:

\begin{equation}
\ell_{ec}=\gamma/G, \quad \ell_{eg} =G/\rho g.
\end{equation}
These are the elastocapillary length $\ell_{ec}$ and the elastogravity length $\ell_{eg}$ respectively. The former describes the cross-over from capillary to elastic behaviour, while the latter indicates when gravity dominates over elasticity. One recovers the usual capillary length of liquid menisci as the geometric mean $\ell_c = \sqrt{\ell_{ec} \ell_{eg}}$. Interestingly, the problem gives rise to characteristic stiffness $G^*$ at which all length scales coincide. Setting $\ell_{ec}=\ell_{eg}$, one finds 

\begin{equation}
G^* =\sqrt{\gamma \rho g}.
\end{equation}
For typical values of $\gamma \sim 10^{-2}$~N/m and $\rho g \sim 10^4$~N/m$^3$, we find this stiffness to be $G^* \sim 10$~Pa. Coincidentally, this stiffness is comparable to the softest hydrogels that can be obtained -- and thus a highly relevant magnitude from an experimental perspective. In fact, the type of meniscus shape is completely determined by the ratio $G/G^*$, which is equivalent to $\sqrt{\ell_{eg}/\ell_{ec}}$.

%
%

\begin{figure*}[t!]
\includegraphics[width=180mm]{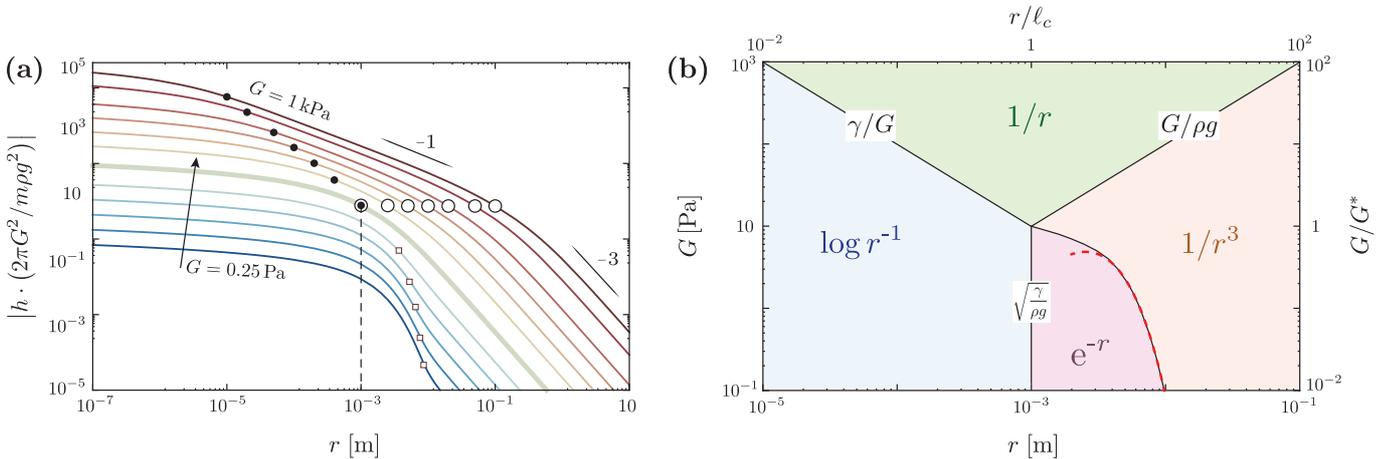}
\caption{\textbf{(a)} Hydrogel shapes $h(r)$ due to indentation by the weight of a particle. Different curves correspond to $G$ in the range of 0.25 Pa - 1 kPa, while $\gamma$ and $\rho g$ are fixed at typical values $10^{-2}$~N/m and $10^4$~N/m$^3$ respectively. For readability of the figure, the vertical axis is scaled in arbitrary manner to avoid overlapping curves. The filled and hollow circles indicate $r=\ell_{ec}=\gamma/G$ and $r=\ell_{eg}=G/\rho g$ respectively, and separate different scaling regimes. At $G=G^{\ast}=10$ Pa (thick green line), these length scales coincide and merge with the capillary length $\ell_{ec}=\ell_{eg}=10^{-3}$ m (black dashed line). The hollow squares represent the transition from capillary to elastogravity meniscus. \textbf{(b)} Phase diagram showing different asymptotic meniscus shapes for varying shear modulus. The dimensionless scales are shown at the right and on top. $\ell_{ec}$, $\ell_{eg}$, and $\ell_c$ define the boundaries of different regimes (black straight lines). The red dashed curve separates capillary and elastogravity meniscus shapes, given by eq.~\eqref{c_eg_boundary}. The associated black curve traces this boundary numerically.}
\vspace{-2mm}
\label{fig2}
\end{figure*}

The key result of this study is shown in Fig.~\ref{fig2}(a), showing the nontrivial meniscus shapes for different stiffnesses. To keep an experimental perspective, we will present the results in dimensional form, with $\gamma = 10^{-2}$~N/m and $\rho g = 10^4$~N/m$^3$, so that $G^*=10$~Pa. The results in Fig.~\ref{fig2}(a) correspond to varying stiffness from the nearly liquid case ($G\approx 0$) to values up to $G=10^3$~Pa. The upper curves correspond to stiff gels. These exhibit three asymptotic regimes, separated by $r \sim \ell_{ec}$ (filled circles) and $r \sim \ell_{eg}$ (open circles). For $r\,\textless\,\ell_{ec}$, the surface resembles a capillary interface where $h \sim \log r$. At distances beyond $\ell_{ec}$ the meniscus follows the classical Boussinesq solution from linear elasticity where $h \sim 1/r$~\cite{johnson_1985}. Finally, as $r$ becomes comparable to $\ell_{eg}$, gravity modifies the deformation and a new asymptotic shape $h \sim 1/r^3$ emerges. This unexpected asymptote reads

\begin{equation}
h \simeq -   \frac{mg\rho g}{\pi G^2} \left(\frac{\ell_{eg}}{r}\right)^3,
\end{equation} 
which we inferred from the limit $\gamma s^2 \rightarrow 0$ for which the inverse of (\ref{hs}) can be found analytically.


When reducing the stiffness, the range over which elastic scaling $\sim 1/r$ is observed gradually diminishes (Fig.~\ref{fig2}(a)). The thick line (green) corresponds to $G^*=10$~Pa, where all lengths coincide at $r=1$~mm, and the naively expected elastic regime has completely disappeared. While much lower $G$ are difficult to realise experimentally, the limit of vanishing stiffness has an intrinsic interest. Namely, the limiting case $G=0$ corresponds to a liquid meniscus. In that case the inverse of \eqref{hs} reduces to the classical solution $-K_0(r/\ell_c)$, which is the zeroth order modified Bessel function of second kind. The liquid meniscus decays as $\e^{-r/\ell_c}$ at large distance. The introduction of a small but finite $G$ modifies this shape at $r\gg\ell_{c}$. The hollow squares of Fig.~\ref{fig2}(a) locate the point where the exponential decay of the liquid meniscus again gives way to the elastogravity scaling of $1/r^3$. These observations provide a strong departure from the previously assumed exponential screening for hydrogels~\cite{Chakrabarti2014,Maha2015}.

To summarise these intricate regimes, we present the various asymptotes in terms of a phase diagram in Fig.~\ref{fig2}(b). The vertical axis indicates the gel's shear modulus $G$, while the horizontal axis is the distance $r$. The relevant dimensionless scales $G/G^*$ and $r/\ell_c$ are indicated on the right and top axis. For substrates with shear modulus of 1 kPa or larger the meniscus is effectively governed by elasticity over the full range of scales. Comparatively softer substrates with $G\sim 100$ Pa exhibit an elasticity dominated region bounded by $\ell_{ec}\sim 100 \, \mu$m and $\ell_{eg}\sim 1$ cm, which shrinks to a point at $G^{\ast}=10$ Pa. As mentioned, the typical shear modulus of extremely soft hydrogels are also around 10 Pa, which makes this region of the phase diagram of particular experimental interest. At this shear modulus, all three length scales are equal, $\ell_{ec}=\ell_{eg}=\ell_c=1$ mm, and the predicted $1/r^3$ elastogravity regimes should be accessible in experiments. As $G\ll G^{\ast}$ the interface mostly resembles a capillary meniscus where $\ell_c$ separates the near-field and far-field behaviour of $\log r$, and $\e^{-r/\ell_c}$ respectively. For completeness, we determine the boundary between capillary and elastogravity by equating the corresponding asymptotes, 

\begin{equation}
\frac{G}{G^*}\simeq \frac{1}{2}\left(r/\ell_c\right)^3 K_0(r/\ell_c).
\label{c_eg_boundary}
\end{equation} 
This is shown as the red dashed line in Fig.~\ref{fig2}(b), in close agreement with the numerical result (solid black curve).

\subsection{Interaction}

With the meniscus shape in hand, we can determine the interaction forces between multiple particles (Fig.~\ref{fig1}). As per~\eqref{f_int}, the interaction force on a particle is directly given by the local slope of $h(\mathbf x)$ at that location. We note that the deformation diverges logarithmically at the location of the particles, which is an artefact of the point-particle approach~\footnote{This paradox of infinite deformation at $\mathbf x_i$ is resolved once the particles are not treated as point-like, but are given a finite size. This changes the shape of the meniscus in the near-field, but does not affect the far-field results.}. This is in direct analogy to (two-dimensional) electrostatics, where the force on a charge is computed by omitting the contribution of the infinite ``self-energy" from the electrostatic potential~\cite{landau1975classical}. Hence, in what follows the force on particle $i$ follows from $h(\mathbf x)$ induced by all \emph{other} particles $j\neq i$.

Here we consider the simple scenario of two identical particles located at $r=0$, and at a distance $r=d$. The interaction force then simply follows from the slope

\begin{equation}
\mathbf F=-mg \d{h}{r}\Bigg|_{r=d}\,\hat {\mathbf r},
\label{f_int_nd}
\end{equation}
where $h(r)$ is the axisymmetric shape derived in the previous section and $\hat {\mathbf r}$ is the radial unit vector. Equation~\eqref{f_int_nd} is equivalent to Nicolson's formula~\cite{nicolson1949} developed in the context of interacting bubbles on a liquid interface. In the present case, for heavy particles on a hydrogel, the slope always remains positive and gives an attractive interaction between the particles. Different meniscus shapes give rise to different interactions laws, and the regimes for the interaction forces are strictly the same as those in Fig.~\ref{fig2}(b). The scaling laws for the forces are simply inferred by taking the derivative of the asymptotes indicated in Fig.~\ref{fig2}(b). For the special case of $G^{\ast}=10$~Pa at distances below $\ell_c\, (10^{-3}$ m), the meniscus shape is given by $\log r$, resulting in $\mathbf F \sim 1/d$, whereas for $d\gg\ell_c$ the elastogravity meniscus leads to an interaction force $\mathbf F \sim 1/d^4$. 

\begin{figure}[t!]
\includegraphics[width=83mm]{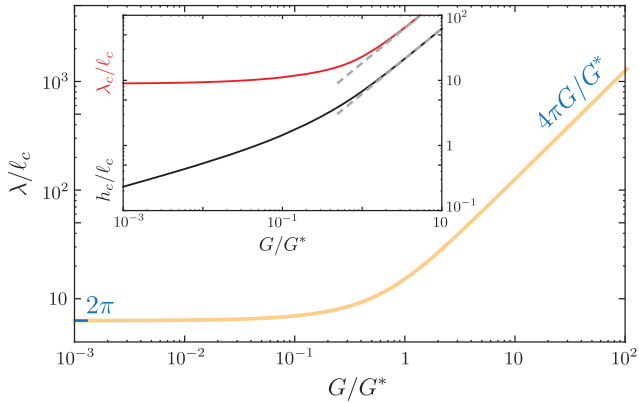}
\caption{Marginal wavelength in Rayleigh-Taylor instability of a hydrogel interface as a function of material stiffness. In the liquid limit we recover $\lambda=2\pi\ell_c$, whereas for an elastic half space we find $\lambda=4\pi\ell_{eg}$. Inset: instability threshold for a thin hydrogel layer showing the variation of critical thickness (black curve) and wavelength (red curve) with shear modulus. The gray dashed lines represent asymptotic relations $h_c\simeq 6.223\ell_{eg}$ and $\lambda_c\simeq 18.445\ell_{eg}$.}
\vspace{-2mm}
\label{fig5}
\end{figure}

\subsection{Instability}

The current formalism also allows to investigate the Rayleigh-Taylor instability (RTI) for hydrogels. Indeed, when a layer of hydrogel in a petri dish is turned upside down the free surface exhibits undulations  \cite{Mora14} that are reminiscent of the RTI of fluid interfaces (e.g. thin viscous films turned upside done). Here we address the RTI for two reasons. First, it serves as a quantitative validation of the additive energy functional (\ref{eq:additive}) for hydrogel interfaces. Second, it allows to unify the liquid and elastic versions of RTI. For fluid layers all wavelengths larger than $\lambda = 2\pi \ell_c$ are unstable~\cite{de2003capillarity}, independently of the layer thickness. For elastic media the situation is more intricate. For an elastic half-space this wavelength reads $\lambda = 4\pi \ell_{eg}$~\cite{Bakhrakh97}, but recent experiments on polyacrylamide hydrogels and theory have identified a threshold sample thickness below which the instability does not occur \cite{Mora14}. Linear stability analysis showed this sample thickness to be $h_0/\ell_{eg} = 6.223$. Our formulation allows to capture all these features in a single, tractable framework.

The onset of instability is studied by plane Fourier waves of wavenumber $q=2\pi/\lambda$. The Fourier transform of (\ref{gov_eq}) then gives

\begin{equation}
\hat{h}(q)={-\hat{\varrho}(q)\over\gamma q^2-\rho g+2G|q|},
\label{hq}
\end{equation}
which is the plane-wave analogue of (\ref{hs}). Importantly, we have flipped the sign of gravity $g\rightarrow -g$ to account for the fact that the hydrogel interface is held upside down. The denominator of the above equation can be interpreted as an \emph{effective stiffness} of the entire layer, as it relates the displacement $\hat h$ to a forcing $\hat \varrho$. The onset of instability is found when this effective stiffness vanishes, as it marks the point where the layer loses its restoring force. Solving for $q$ where the denominator of (\ref{hq}) vanishes, we find the wavelength

\begin{equation}
\lambda/\ell_c=2\pi\left({G\over G^{\ast}}+\sqrt{1+\left({G\over G^{\ast}}\right)^2}\right).
\label{l_int}
\end{equation} 
Figure~\ref{fig5} shows how this wavelength evolves with $G/G^{\ast}$. It indeed bridges between the liquid and elastic RTI. For a liquid interface ($G/G^* \rightarrow 0$) we recover $\lambda = 2\pi\ell_c$, while the elastic half space ($G/G^*\rightarrow\infty$) is unstable under gravity for perturbations with wavelengths larger than $\lambda = 4\pi G/G^{\ast} \ell_c=4\pi\ell_{eg}$. 

The final step is to incorporate the effect of a finite layer thickness $h_0$, and see why a threshold appears for hydrogels but not for liquid layers. For this case, we use a modified Green's function such that $\hat{\sigma}_{e\ell}(q)=G q^2 \hat{h}(q) \hat{K}(q) \rightarrow G \hat{h}(q) k(\bar q)/h_0$. Here we introduced the dimensionless wavenumber $\bar q = q h_0$, and the Green's function for a layer attached to a rigid base in dimensionless form \cite{Xu14964}

\begin{equation}
k(\bar q)=2 \bar q\left[{\cosh(2\bar q)+2(\bar q)^2+1\over\sinh(2\bar q)-2\bar q}\right].
\label{k_ft}
\end{equation}
In the limit of $\bar q \rightarrow\infty$,  one has $k(\bar q)=2|\bar q|$ and we recover the half space result of~\eqref{hq}. For finite thickness, the condition of a vanishing denominator of the modified (\ref{hq}) reads

\begin{equation}
\left({h_0\over\ell_c}\right)^2-{G\over G^{\ast}}k(\bar{q})\left({h_0\over\ell_c}\right)-\bar{q}^2=0,
\label{h_cr_in}
\end{equation}
which marks the onset of the Rayleigh-Taylor instability. For any value $G/G^* \neq 0$, this equation indeed predicts a minimum layer thickness below which the interface remains stable against perturbations of any wavenumber, thus confirming the existence of an instability threshold. The inset of Fig.~\ref{fig5} shows this critical thickness $h_c$ and the corresponding critical wavelength ($\lambda_c$) as a function of $G/G^*$. For stiffer hydrogels, the third term in~\eqref{h_cr_in} can be neglected and $h_c$ is simply given by the minimum value of $k(\bar{q}_c)\simeq 6.223$ as $h_c=6.223\ell_{eg}$. Hence, we perfectly recover the threshold obtained by Mora \emph{et al.}~\cite{Mora14}, who solved the bulk elastic equations including gravity as a bulk force. It confirms the validity of using an additive energy functional (\ref{eq:additive}). However, we can now investigate what happens when the hydrogels become softer and consider (\ref{h_cr_in}) over the full range of $G/G^*$. The result is shown in the inset of Fig.~\ref{fig5}. In the limit of $G/G^*\rightarrow 0$, we find $h_c/\ell_c\sim(G/G^{\ast})^{1/3}$, so that indeed the threshold vanishes in the liquid limit.

\section{Discussion}

In summary, we have computed how hydrogels deform under the influence of particles and how this leads to mutual interactions similar to the Cheerios effect. It is shown that both surface tension and gravity (hydrostatic pressure) can play a role for sufficiently soft materials. This leads to a variety of regimes, which were classified in detail (Fig.~\ref{fig2}b). Importantly, we identified a characteristic shear modulus $G^* = \sqrt{\gamma \rho g}$, which for real materials is typically a few tens of Pascals. A hydrogel's mechanical response is solid-like for $G \gg G^*$, but becomes more liquid-like when $G \sim G^*$. The ratio $G/G^*$ also governs the nature of the Rayleigh-Taylor instability for an inverted layer of hydrogel.

The role of both gravity and surface tension was previously appreciated for experiments on the ``elastic Cheerios effect", where spheres and cylinders on a hydrogel were indeed found to attract \cite{Maha2015, Chakrabarti2014}. While qualitatively consistent with our findings, these studies postulated that the force of interaction decays exponentially. This was inspired by the interactions on a purely liquid interface, and a ``modified" capillary length was introduced to account for elasticity. However, our calculations reveal a different picture, since elasticity changes the decay from exponential to algebraic. It would be important to  validate these observations experimentally. An interesting extension of the present study is to consider very small particles, for which the adhesion to the gel dominates over the particle weight. This situation closely resembles that of liquid drops, for which the loading is tensile at the contact line and compressive in the contact zone. In Appendix B we show that the interaction force on a ``sticky" particle reads $\mathbf F\sim-\nabla \nabla^2h$, as opposed to $\sim \nabla h$ for ``heavy" particles. 

From a more general perspective, similar elasto-gravity problems are encountered in geological contexts such as vulcano deformations \cite{barbot10}. We therefore expect that the presented energy approach, and the explicit elasto-gravity functional, will serve for a variety of problems.

{\bf Acknowledgments.} 
We thank B. Andreotti, L. Botto, M. Costalonga and L. van Wijngaarden for discussions. AP and JHS acknowledge financial support from ERC (the European Research Council) Consolidator Grant No. 616918.

\section{\label{sec:energy}Appendix A: Elastic Energy}

In linear elasticity the strain energy (in absence of shear traction) can be written as a surface integral

\begin{equation}\label{eq:enelastic}
E_{\rm el} = \frac{1}{2} \int d\mathbf x \, \sigma(\mathbf x)  h(\mathbf x).
\end{equation}
To obtain $E_{\rm el}$ entirely in terms of $h(\mathbf x)$, we need to replace the normal stress $\sigma$. In linear elasticity, the deformation can be expressed as a convolution of $\sigma$ with the Green's function $K(\mathbf x)=1/|\mathbf x|$:

\begin{eqnarray}\label{eq:3D}
h(\mathbf x) &=&  \frac{1}{4\pi G} \int d\mathbf x' \, \frac{\sigma(\mathbf x')}{|\mathbf x -\mathbf x'|},
\end{eqnarray} 
or 

\begin{equation}
\nabla h(\mathbf x)=-{1\over 4\pi G}\int d\mathbf{x'}\,\sigma(\mathbf x')\,{(\mathbf x -\mathbf x')\over|\mathbf x -\mathbf x'|^3}.
\label{}
\end{equation} 
This is in the form of a two-dimensional Hilbert transform~\cite{duffin57}, which has as its inverse
 
\begin{equation}
\sigma(\mathbf x)={G\over\pi}\int d\mathbf{x'}\,\nabla h(\mathbf x')\cdot{(\mathbf x -\mathbf x')\over|\mathbf x -\mathbf x'|^3}.
\label{}
\end{equation}
So, the energy functional becomes

\begin{equation}
E_{\rm el}[h]={G\over2\pi}\int d\mathbf{x}\, h(\mathbf x)\int d\mathbf{x'}\,\nabla h(\mathbf x')\cdot{(\mathbf x -\mathbf x')\over|\mathbf x -\mathbf x'|^3}.
\label{}
\end{equation}
Integration by parts then gives~\eqref{E_el}.

\section{Appendix B: Adhesive particles}

Adhesive particles without weight can be modelled as axisymmetric distributions of normal traction $T_i$ centered around the particle position $\mathbf x_i$. As the particles are weightless and axisymmetric, the zeroth and first moments vanish, i.e.  $\int \id\mathbf x \,  T_i(\mathbf x)=0$, and $\int \id\mathbf x \,  T_i(\mathbf x)\mathbf x=0$. The work done by this traction gives the coupling

\begin{equation}
E_{\rm adh} =  \sum_i \int \id\mathbf x \, T_i(\mathbf x - \mathbf x_i) h(\mathbf x)\simeq  \sum_i q_i\nabla^2 h(\mathbf x_i).
\label{e_adh}
\end{equation}
Here $q_i=\int \id\mathbf x \,  T_i(\mathbf x) |\mathbf x|^2$ is the quadrupole moment. The second step in (\ref{e_adh}) is obtained by expanding the field $h(\mathbf x)$ around $\mathbf x_i$ as

\begin{equation}
h(\mathbf x)  = h(\mathbf x_i) +  (\mathbf x - \mathbf x_i)\cdot \nabla h(\mathbf x_i)  
+ \frac{1}{2} (\mathbf x - \mathbf x_i)^T \cdot  \mathbf H(\mathbf x_i) \cdot (\mathbf x - \mathbf x_i),
\end{equation}
(where $\mathbf H$ is the Hessian). Inserted in $E_{\rm adh}$, and using the vanishing zeroth and first moments, indeed gives

\begin{eqnarray}
\int \id\mathbf x \, T_i(\mathbf x - \mathbf x_i) h(\mathbf x) &\simeq &
\nabla^2 h(\mathbf x_i) \int \id\mathbf x \,  T_i(\mathbf x) |\mathbf x|^2. 
\end{eqnarray} 
The total energy is $E = E_{\rm int}[h] + E_{\rm adh}$, so that for adhesive particles we find $\mathbf F_i = - \frac{\partial E}{\partial \mathbf x_i} = - q_i \nabla \nabla^2 h$. The field $h(\mathbf x)$ is still found by solving 
$\sigma + \varrho =0$, but now the loading becomes $\varrho(\mathbf x) =   \sum_i T_i \left( \mathbf x - \mathbf x_i \right)$.

\end{document}